# A manager´s view on large scale XP projects

**Bernhard Rumpe**
Software & Systems Engineering
Munich University of Technology
Arcisstr. 21
D-80333 Munich, Germany
+49 89 289 28129
Bernhard.Rumpe@in.tum.de

**Peter Scholz**
Department of Computer Science
University of Applied Sciences Landshut
Am Lurzenhof 1
D-84036 Landshut, Germany
+49 871 506 679
Peter.Scholz@fh-landshut.de

**ABSTRACT**
XP is a code-oriented, light-weight software engineering methodology, suited merely for small-sized teams who develop software that relies on vague or rapidly changing requirements. Being very code-oriented, the discipline of systems engineering knows it as approach of incremental system change. In this contribution, we discuss the enhanced version of a concept on how to extend XP on large scale projects with hundreds of software engineers and programmers, respectively. A previous version was already presented in [1]. The basic idea is to apply the "hierarchical approach", a management principle of reorganizing companies, as well as well-known moderation principles to XP project organization. We show similarities between software engineering methods and company reorganization processes and discuss how the elements of the hierarchical approach can improve XP. We provide guidelines on how to scale up XP to very large projects e.g. those common in telecommunication industry and IT technology consultancy firms by using moderation techniques.

## 1  INTRODUCTION

Extreme Programming (XP) [2,11] is the most prominent of the new generation of light-weight (also called agile) methodologies for small-sized teams developing software with vague or rapidly changing requirements. XP can be regarded as an explicit reaction to the complexity of today's modelling techniques like the Unified Process [3], the V-model [4], Catalysis [5], or the Open Modeling Language [6]. XP focuses on a system of best practices that are deeply interconnected, disregarding many others used by other methodologies. XP will evolve, reducing its weaknesses and increasing its strengths. This article suggests an improvement in one of its obvious weaknesses:

XP is designed for a single small team of less than a dozen team members. Therefore, it has its problems to scale up for larger projects. In those cases, direct team communication is no longer possible without any additional support. Fortunately, applying the XP approach in projects seems to considerably downsize the number of necessary participants, but there is still a number of areas, where hundreds of developers work on producing one single software product. For example, the telecommunication industry is under enormous pressure to add and improve functionality of their products. The time to market span in the mobile phone business needs fast and flexible process for large projects. Switching systems need to be adapted for each customer and for each country: XP is just starting to play its role here too. The main obstacles against scaling up of XP are lack of documentation (therefore the exponential increase of necessary communication between developers), lack of stable interfaces and stable requirements. Consequently, scaling up of XP will probably be indispensable in order to adopt methodical practices from other methodologies.

From the discipline of systems engineering, we are acquainted with three approaches suited to manage a reorganization project. In the Total Systems Approach [7], the desired properties of a new system are first defined and then the whole system is introduced into the new organization like a big-bang invention. In the Incremental Systems Approach, a set of small changes incrementally leads to local optimisation. Through small changes of the company structure and organization and its supporting software system, a series of small localized improvements lead to a sub-optimal organization form. As both approaches have several drawbacks, discussed below, system engineering provides a third approach called "Hierarchical structuring" of system development. This approach combines the advantages of both other approaches, usually leads to better reorganization projects, and therefore provides an overall optimisation.

In [1] similarities between the extreme programming approach and the incremental systems developing approach have been discussed and the combination of these two approaches from systems engineering has been transferred to the software engineering discipline. There are two basic advantages: (1) the combination leads to a scale up of the extreme programming approach to larger projects by hierarchically structuring the teams, and (2) it features a successful methodology for the organization of the hierarchical approach which can be transferred to the software engineering discipline. Both points have interesting aspects. Of course, scaling XP up to a larger project allows to apply the XP approach even if the system becomes more complex needing more people to be involved. Another advantage is that there is a proven methodology to get a hierarchical reorganization process organized; this can be adopted by the software engineering discipline.

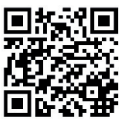

This contribution is structured as follows. In Section 2 we introduce the new approach to reorganize parts of the company, discuss the analogy to software process models and point out some improvements for XP. In Section 3 we discuss management techniques specifically suited to supplement the new hierarchical XP approach. A brief overview of the aspects of XP which are of interest in this context can for instance be found in [1].

## 2 HIERARCHICAL XP

Today it is for all companies imperative to supply their business with extensive software support. A company reorganization always goes together with the adaptation of existing and the introduction of new software and all too often also the introduction of new software does or should go along with adaptation of the companies business processes and structures. Therefore, it is a natural consequence to combine suitable approaches that come from technical and management disciplines. In hierarchical XP, two approaches with similar characteristics are combined. The holistic approach (from systems theory) has several characteristics in common with the classical software engineering approaches, starting with the Waterfall model, but also newer object-oriented approaches, like the Unified Process [3]. They e.g. share a centralized approach providing a small coordination team with great power, but lack adequate customer/employee participation.

The incremental approach (from systems theory) compares well to XP. Both are rather decentralized and both focus minor on local improvements of existing structures/systems. Such improvements can be released early and get a fast feedback. Their major advantages are: high involvement of employees/customers, and as a result, high acceptance of the solution. XP and the incremental approach do have also some disadvantages in common: (1) applying this approach to several local problems does usually not lead to a shared improvement with multiple teams. Instead, local improvements may contradict each other, (2) the approach is unstructured and can therefore not be used for working out an overall concept by a complex problem where an involvement of several persons is necessary.

In systems theory, the hierarchical approach was developed as a combination of the holistic and incremental approaches and has been carried over to the software development discipline in [1]. This approach starts from the extreme programming approach and builds a hierarchical structure upon it. The advantages of this approach have partly been discussed before: While largely retaining a light-weight methodology, it becomes feasible to structure larger projects into a bunch of smaller XP projects that still have a common target to achieve. The approach basically consists of two important elements: (1) on the top-level we set up a goal-oriented project management (called steering committee) that organizes the problem as a high-level structure by working out a rough concept, (2) each of the now localized problem parts is solved in an extreme programming approach by its own XP team. The following figure demonstrates this advantage of the hierarchical approach compared to the other two approaches. Each circle is a team member, tight connections of circles form a team.

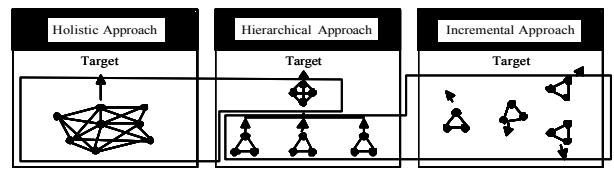

**Figure 1: Comparison of three approaches**

The XP teams function primarily on an independent basis; nevertheless, they are coupled by a top-level management team, called "steering committee", that keeps track on the overall goal and measures local improvements. It is important to keep arising cross-dependencies as lean as possible. However, the complexity of today's information systems, at least partly arises from the still insufficient mechanisms to define crisp and small interfaces between software parts. Dynamic restructuring of the XP teams is useful to flexibly react on varying workloads. So over time, the project structure e.g. splits as follows:

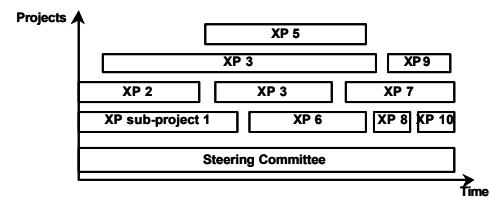

**Figure 2: Hierarchic project structure**

By organizing the software development process in a hierarchical manner the focus is given on one common target and a structured process to reach this goal is used. The involvement of the employees will lead to a high acceptance for the solution. Ideally XP project teams are defined in a similar way as company departments are. A certain part of the software infrastructure of a company is not localized in one (or a few) departments, but its usage spread over a number of departments. This can e.g. be handled by identifying pilot departments that are able to cover the needs and desires of other departments' users as well.

These considerations show that the hierarchical extreme programming approach needs a focussed, yet lean project organization. Five major principles can be identified that characterize the hierarchical approach:

1. Customer participation: the solution is worked out with the customer/employee to reach a high acceptance. This is in particular important for the customers to accept the resulting new software system/company structure.
2. The whole system is divided up into subsystems with a lean and crisp interface. The inputs and outputs, namely the data structures and the information flow between the subsystems need to be clearly defined. Subsystems are implemented respectively evolved through XP teams.
3. Each XP team targets its associated subsystem, thus contributing to the main target, namely the development of the whole system.
4. The worked out software solutions will be improved like in an incremental process to be successful very quickly. In a number of releases the team explores and extends the desired system functionality.
5. The hierarchical approach is well organized with a project team and a steering committee. The steering committee is an ideal place to develop and maintain the common system goals.

Most of the additional principles and practices of XP, that have been introduced in Section 2, carry over to the hierarchical approach without major changes. However, some of these principles need slight enhancement. Automated test suites become even more important when the XP teams are connected though interfaces. Specific test suites check functionality against mock interfaces. Further tests need to check the correctness of the cross project functionality and therefore the correctness of the interfaces.

## 3 PROJECT MANAGEMENT FOR LARGE SCALE XP PROJECTS

As we have seen, hierarchic XP needs a focused set of project management techniques to handle the issues arising in the steering committee. In this section, we will motivate how project management, enriched by additional moderation and communication techniques can be applied to this kind of large scale XP software development projects. Recall the four values of XP: two of them have been "communication" and "early feedback". Hence, the success of every XP project very much depends on how these two values are reached. Both communication and feedback become harder to realize with every single additional team member involved in the team. Hence, we have to find solutions on how to guarantee efficient communication and feedback for every size of XP project.

In general, in large scale projects, i.e. in projects where many persons are involved in, the communication overhead increases dramatically. This context was first shown by Brooks. Brooks's law, pictured in Figure 3, outlines the relationship between the number of persons involved in a project and the time-to-product.

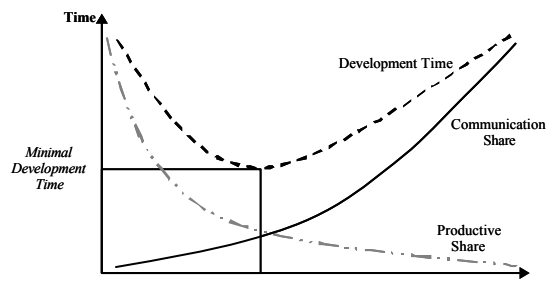

**Figure 3: Brooks's law**

Brooks's law shows that the time-to-product cannot be decreased below a certain point by just adding project members. If this number exceeds a certain point, the communication overhead takes over. This increasing communication overheads limits the optimal number of project team members to a certain number. In large scale XP projects, however, a lot more developers than this optimal number are involved. Hence, measures have to be taken that allow for additionally increasing the number of team members. The first and one of the most efficient lever to do so is to split the team into XP subteams as done in hierarchical XP. For the still necessary communication moderation techniques are applied.

Moderation in XP projects aims at involving all project members as efficiently as possible in all project phases. This ensures that the members' ideas and energies can be bundled up and therefore optimally brought into the project. As a consequence, all team members pull together. However, to be effective, moderation has to be carried out in a systematic, structured, and open manner, that is, without any manipulation of any kind (for instance, by political top management conflicts). Project work that is guided this way by a professional moderator makes a lot of fun. In addition, moderation causes a number of further advantages: (1) all project members are concentrated on the working content, only, (2) all results get transparent, (3) the cooperation, team spirit and therefore, the overall company culture improves, and (4) the motivation of each XP project member increases.

**What is the moderator's task in XP projects?**
She or he has to support the programmers in a way that problems can be solved by themselves, i.e. by team work. Also, the efficiency with respect to the project return on investment has to be increased by the moderation. Furthermore, solution concepts should be worked out that are accepted by both programmers and the top management. Though the moderator has a strategic position within the project he or she also gives know-how to the project members. However, know-how in this context equally means strategic, technical and application know-how. A good XP moderator knows all moderation methods (the moderation tool box) and has understood the XP idea.

The larger the project the less likely it is that the moderator will programme himself and therefore has not to be a good programmer. The moderator must be like an "obstetrician" so that complex ideas can be born, formulated, cut into components for the subprojects and realized. Finally, he or she takes care that the potential of each project member can be exhausted in an optimal way. Altogether, the moderator supports the team in questions of method, motivation, communication, and cooperation. As in the hierarchical approach, the subteams flexibly reorganize during the project, he also holds a part of the responsibility to enable appropriate reorganization, but should not be the finally responsible person.

**How can the concept of moderation be applied to very large XP projects?**
In order to work efficiently, each moderator merely is able to support up to 6-8 pairs of programmers, which to our experience means an average of three XP subprojects. The interesting questions is, how to apply moderation to XP projects with 100 and more developers.

The idea is that larger teams consisting of 6-8 pairs of programmers is coached by its "own" moderator, smaller teams share moderators. Every moderator is responsible for the knowledge transfer within his team, as he is also member of the steering committee. This enables in addition to the intra-team knowledge transfer also an inter-team knowledge transfer. At least the moderators of the steering committee meet regularly by establishing "heures fixe" (like the well-known "jour fixe" but just carried out in a higher frequence): All moderators involved in a particular large scale XP projects meet each other daily either in the morning, or in the noon, or in afternoon hours to exchange project knowledge from their teams.

In addition it is feasible to support the teams of programmers by a team of developers who are responsible for unit testing. This team is responsible for the overall function tests with particular focus on correct handling on the interfaces in both directions. Furthermore, development team and unit testing team should meet altogether about every four weeks in order to identify, discuss, and solve development, quality, or process problems. These meetings also should provide a platform for know-how exchange.

There are no additional, disciplinary organization structures. This way, the organization is kept simple, flat, and therefore powerful.

## 4 CONCLUSION

This paper focuses on the particular question how to optimize and reorganize companies that make heavy use of software products. First, hierarchical XP is introduced as a software engineering method for large scale projects that possible structures its sub-projects along business or organizational structures. If a company is reorganized, this usually means restructuring its software products, its databases and network infrastructure, because a company reorganization usually concentrates on optimization of its business cases. In this paper, we have extended the extreme programming approach by elements of the hierarchical reorganization process leads to a considerable scale-up of the XP approach. Furthermore, the XP approach extended this way can nicely be integrated with the hierarchical reorganization process allowing the use of both at the same time. Furthermore, we have identified and discussed moderation techniques that are used to coach XP development teams.

**ACKNOWLEDGEMENTS**
This work was partially supported by the Bayerisches Staatsministerium für Wissenschaft, Forschung und Kunst through the Bavarian Habilitation Fellowship, the German Bundesministerium für Bildung und Forschung through the Virtual Software Engineering Competence Center (ViSEK) and Siemens.